\documentclass{article}
\usepackage[left=1in, right=1in, top=1.5in, bottom=1.5in, paper=letterpaper]{geometry}
\usepackage{amssymb}
\usepackage{amsmath}
\usepackage{graphicx}
\usepackage{multirow}
\usepackage{hhline}
\usepackage[titletoc,title]{appendix}
\usepackage[table]{xcolor}
\usepackage{listings}
\usepackage{subfig}
\usepackage{framed}

\definecolor{codegreen}{rgb}{0,0.6,0}
\definecolor{codegray}{rgb}{0.5,0.5,0.5}
\definecolor{codepurple}{rgb}{0.58,0,0.82}
\definecolor{backcolour}{rgb}{0.95,0.95,0.92}

\lstdefinestyle{mystyle}{
    backgroundcolor=\color{backcolour},
    commentstyle=\color{codegreen},
    keywordstyle=\color{magenta},
    stringstyle=\color{codepurple},
    basicstyle=\footnotesize,
    breakatwhitespace=false,
    breaklines=true,
    captionpos=b,
    keepspaces=true,
    showspaces=false,
    showstringspaces=false,
    showtabs=false,
    tabsize=2
}

\usepackage{hyperref}
\hypersetup{
	colorlinks   = true,
	citecolor    = blue,
	linkcolor    = blue,
}
\usepackage{url}
\urldef{\mailsa}\path|sbadihi@ce.sharif.edu|
\urldef{\mailsb}\path|, heydarnoori@sharif.edu|

\usepackage[colorinlistoftodos,prependcaption,textsize=tiny]{todonotes}
\newcommand{\unsure}[2][1=]{\todo[linecolor=red,backgroundcolor=red!25,bordercolor=red]}
\newcommand{\change}[1][1=]{\todo[linecolor=blue,backgroundcolor=blue!25,bordercolor=blue]}
\newcommand{\info}[2][1=]{\todo[linecolor=OliveGreen,backgroundcolor=OliveGreen!25,bordercolor=OliveGreen]}
\newcommand{\improvement}[2][1=]{\todo[linecolor=Plum,backgroundcolor=Plum!25,bordercolor=Plum]}
\newcommand{\thiswillnotshow}[2][1=]{\todo[disable]}

 \newcommand{\comment}[1]{}

\newcommand{\secref}[1]{Section~\ref{sec:#1}}
\newcommand{\tabref}[1]{Table~\ref{tab:#1}}
\newcommand{\figref}[1]{Figure~\ref{fig:#1}}

\newcommand{\eqaref}[1]{Equation~\eqref{eq:#1}}

\begin{document}

%\title{On the Use of Crowd in Generating Natural Language Summaries}
\title{Generating Code Summaries Using the Power of the Crowd}

\author{Sahar Badihi \and Abbas Heydarnoori}

\maketitle

%%%%%%%%%%%%%%%%%%%%%%%%%%%%%%%%%%%%%%%%%%%%%%%%%%
%%
%%                     ABSTRACT
%%
%%%%%%%%%%%%%%%%%%%%%%%%%%%%%%%%%%%%%%%%%%%%%%%%%%
%% Should be less than 150 words.

\begin{abstract}
One of the first steps to perform most of the software maintenance activities, such as updating features or fixing bugs, is to have a relatively good understanding of the program's source code which is often written by other developers. A code summary is a description about a program's entities (e.g., its methods) which helps developers have a better comprehension of the code in a shorter period of time. However, generating code summaries can be a challenging task. To mitigate this problem, in this article, we introduce \emph{CrowdSummarizer}, a code summarization platform that benefits from the concepts of crowdsourcing, gamification, and natural language processing to automatically generate a high level summary for the methods of a Java program. We have implemented CrowdSummarizer as an Eclipse plugin together with a web-based code summarization game that can be played by the crowd. The results of two empirical studies that evaluate the applicability of the approach and the quality of generated summaries indicate that CrowdSummarizer is effective in generating quality results.
\end{abstract}

%%%%%%%%%%%%%%%%%%%%%%%%%%%%%%%%%%%%%%%%%%%
%%%
%%%             Introduction
%%%
%%%%%%%%%%%%%%%%%%%%%%%%%%%%%%%%%%%%%%%%%%%
\section{Introduction}
To answer users changing requirements, software systems must continuously evolve~\cite{lehman}. Having developers to understand a program's source code which may include thousands of lines of code, and focus on those parts on which they want to perform their maintenance activities, is the main reason of 60\% to 90\% of the overall software maintenance and evolution costs~\cite{salton-automatic}. This is mainly due to the fact that original developers of the software system may no longer be available in the team, and thus, any maintenance tasks in this phase can be highly error prone. Hence, program comprehension and understanding the rationale behind the code are the most critical and time consuming steps for developers during software maintenance and evolution~\cite{von}, being that they spend substantial efforts on reading and finding relative parts of the code rather than applying the modification~\cite{ko}. However, using \emph{code summaries} can be a solution to this difficulty~\cite{McBurney:TSE:2016}. A code summary is a brief description about the functionality and the purpose of a section of the source code. Code summaries can help programmers find the relevant parts of the source code to their maintenance tasks much easier and faster~\cite{McBurney:TSE:2016, haiduc-use}. McBurney et~al.~\cite{mcburney-empirical} pointed out that the code authors use more development details and low level implementation informations. Meanwhile, readers are those who want to understand the code; therefore, they must be able to deduce the
concepts from the low level details. Ignoring the way which a reader would write a summary is the most important defect in the existing source code summarization approaches.

\emph{Crowdsourcing} is a novel problem-solving approach which is a way to outsource different tasks to a crowd of people through open call (e.g., via the Internet) instead of traditional suppliers~\cite{howe-rise}. In recent years, crowdsourcing has attracted significant attentions to support a wide range of software engineering activities like requirements engineering, design, coding, testing, and evolution and maintenance~\cite{mao-survey}. However, to the bests of our knowledge, using a crowd of developers to write summaries for a piece of a code is a novel idea in the context of program comprehension which we will present in this article. More specifically, here we explain our crowdsourcing platform, named \emph{CrowdSummarizer}, that applies the concepts of crowdsourcing, gamification, and natural language processing to motivate developers to write high level summaries for the methods of a Java program. Furthermore, CrowdSummarizer continuously learns a set of weights and sentence templates from the set of methods and their corresponding summaries which have been collected from the crowd so far, and uses them to also automatically generate the summaries. This helps developers to save time and not necessarily wait for the crowd to write summaries for their methods.

We have implemented CrowdSummarizer as a web-based game as well as an Eclipse plugin that not only automatically generates natural language summaries for the methods of a Java program, but also submits those methods to CrowdSummarizer's website for the summarization by the crowd. We have used this implementation in an empirical study with 149 developers of various levels of experience to generate summaries for 128 methods with different properties from 11 widely used open-source Java applications. The results of the study showed that CrowdSummarizer is applicable for developers to use it in practice. In addition, the results of another empirical study performed with the help of 14 experts illustrated that CrowdSummarizer can automatically generate accurate and comprehensible summaries.

Our main contributions of this article include:
\begin{enumerate}
\item Presenting a novel approach for generating natural language summaries for a Java program and its methods using a topic modelling approach and power of the crowd.
\item Developing a crowdsourcing platform which uses the gamification elements to encourage users' engagement.
\item Evaluation of the applicability and usefulness of CrowdSummarizer platform as well as the quality of our automatically generated summaries.
\item A complete implementation of our approach for Java language as an Eclipse plug-in.
\end{enumerate}
The remainder of this article is organized as follows.
We ﬁrst provide a motivating example, used through-
out this article (Section \ref{motivating-example}). We then introduce the details of CrowdSummarizer technique (Section \ref{proposed_approach}) and its implementation (Section \ref{sec:impl}). We describe an example which CrowdSummarizer try to solve it (Section \ref{exp}). Next, we present two empirical evaluation to evaluate the applicability of the CrowdSummarizer platform, and different aspects of summaries generated for the 78 Java methods from the 11 open-source Java applications (Section \ref{sec:eval}). Finally, we discuss several observations on collected summaries from the CrowdSummarizer (Section \ref{discussion}); compare the technique
with related work (Section \ref{sec:relatedwork}); and conclude (Section \ref{sec:conclusion}).
%%%%%%%%%%%%%%%%%%%%%%%%%%%%%%%%%%%%%%%%%%%
%%%
%%%            Motivating Example
%%%
%%%%%%%%%%%%%%%%%%%%%%%%%%%%%%%%%%%%%%%%%%%
\section{Motivating Example\label{motivating-example}}
We clarify the problem that CrowdSummarizer aims to tackle using an example. Suppose a novice developer has been asked to apply changes to a program which she has no prior background about its source code. For instance, suppose she has been asked to add the following new feature to the \emph{JEdit}\footnote{\url{http://www.jedit.org/}} editor which is currently 117 KLOC, and includes 555 Java files and 7161 methods: \emph{``support for advanced mouse buttons and mouse shortcuts"} (feature request \#476\footnote{\url{https://sourceforge.net/p/jedit/feature-requests/476/}}).

\begin{figure}[t]
  \centering
  \includegraphics[width=0.75\textwidth]{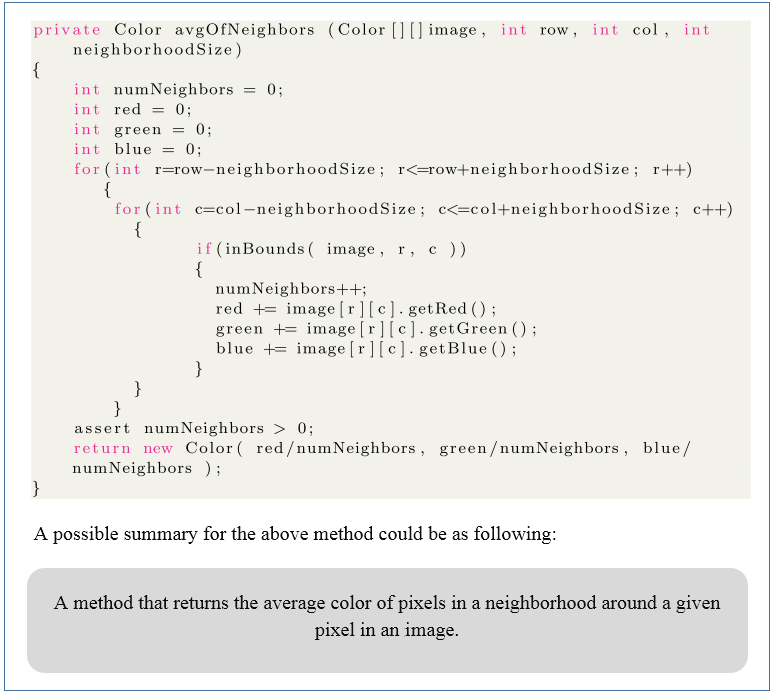}
  \caption{An example code snippet and its sample summary.}\label{fig:example}
\end{figure}

The first step to make the above change is to have a relatively good understanding of the program's source code which is written by other developers who may have left the team. The source code documentation is often one of the possible solutions which can help mitigate the burden of this problem. However, it is often missing or outdated, and also time consuming to read. Reading the entire source code in detail or skimming through it are two other extreme approaches for comprehending the program. Nevertheless, they are time consuming and imprecise, respectively~\cite{haiduc-use}. Code summarization is a middle tactic that mediates between the two tactics by generating high level descriptions about the elements of a program, and can result in a better understanding of the code in a shorter period of time~\cite{haiduc-use}. \figref{example} represents a sample code snippet together with its summary.

We address the issue of source code summarization as following:

{\it{Given a Java program, automatically discover the list of latent topics in the program, classify the methods of the the program into these topics and generate descriptions about relations between the topics and the program's methods.}}

Specifically, the problem of method summarization:

{\it{Given the signature and body for the method $m$, automatically generate sentences describe the high level action of the method from readers' (i.e. a person who didn't write the code) perspective.}}
%%%%%%%%%%%%%%%%%%%%%%%%%%%%%%%%%%%%%%%%%%%
%%%
%%%          Proposed Approach
%%%
%%%%%%%%%%%%%%%%%%%%%%%%%%%%%%%%%%%%%%%%%%%
\section{Proposed Approach\label{proposed_approach}}
\begin{figure}[h]
  \centering
  \includegraphics[width=\textwidth]{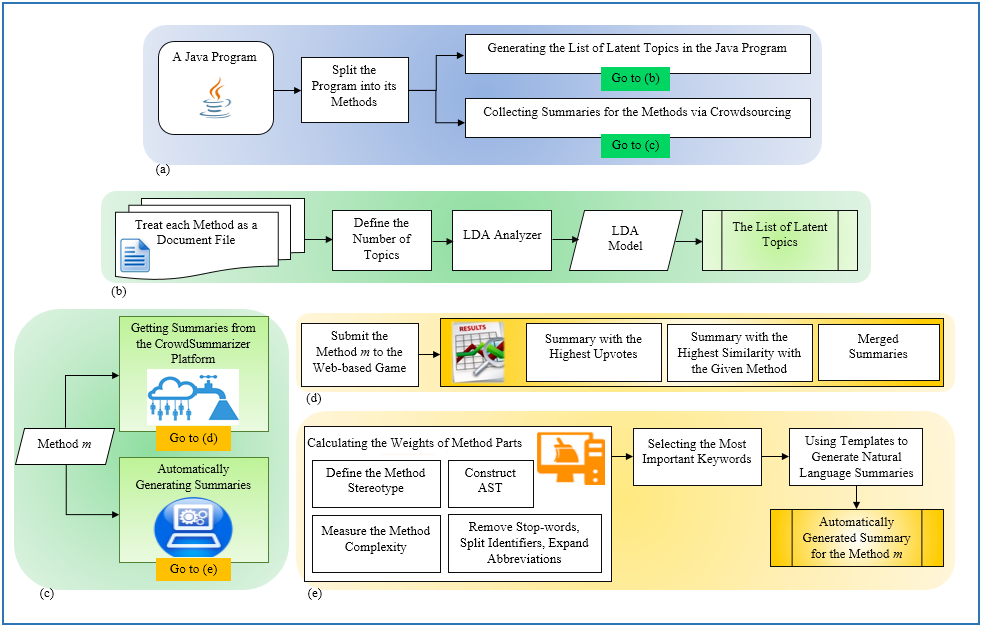}
  \caption{The overall process of CrowdSummarizer and its components: (a) The overall process, (b) Generating the list of latent topics in a Java program, (c) Collecting summaries for the methods via crowdsourcing, (d) Getting summaries from the CrowdSummarizer platform, (e) Automatically generating summaries.}\label{fig:overall-process}
\end{figure}

Before delving into the details of CrowdSummarizer, we first provide an overview of the approach. As can be seen in \figref{overall-process}, the proposed approach has two main components to automatically generate summaries for a Java program: (i) the \emph{LDA} (\emph{Latent Dirichlet Allocation}) component to extract latent topics in a Java program, and to classify its methods into those topics (\figref{overall-process}(b)), and (ii) the crowdsourcing component to generate natural language summaries for each method (\figref{overall-process}(c)). The crowdsourcing component, which is the main focus of this article, constitutes two subcomponents. The first subcomponent (\figref{overall-process}(d)) utilizes a web-based game to get the summaries for each method from the crowd and to generate its results (the \emph{active} use of the approach). Meanwhile, the second subcomponent (\figref{overall-process}(e)) generates automated natural language summaries for each method by using the information which has been gained so far from the naming conventions and the linguistic knowledge as well as the summaries derived from the crowd (the \emph{passive} use of the approach).

%%%%%%%%%%%%%%%%%%%%%%%%%%%%%%%%%%%%%%%%%%%
%%%
%%%  Generating the List of Latent Topics in a Java Program
%%%
%%%%%%%%%%%%%%%%%%%%%%%%%%%%%%%%%%%%%%%%%%%

\subsection{Generating the List of Latent Topics in a Java Program}\label{sec:topic-model}
The aim of this phase is to automatically generate the list of latent topics in a Java program, and to classify the methods of that program into those topics. This would help developers to get a high level view about that program and the distribution of the methods to find out which methods work on a same concept (i.e. topics). For this purpose, we apply the LDA technique~\cite{blei2003latent} as the topic modelling approach to extract topics.
\subsubsection{Latent Dirichlet Allocation\label{pat:LDA}}
\begin{figure}[h]
  \centering
  \includegraphics[width=300px]{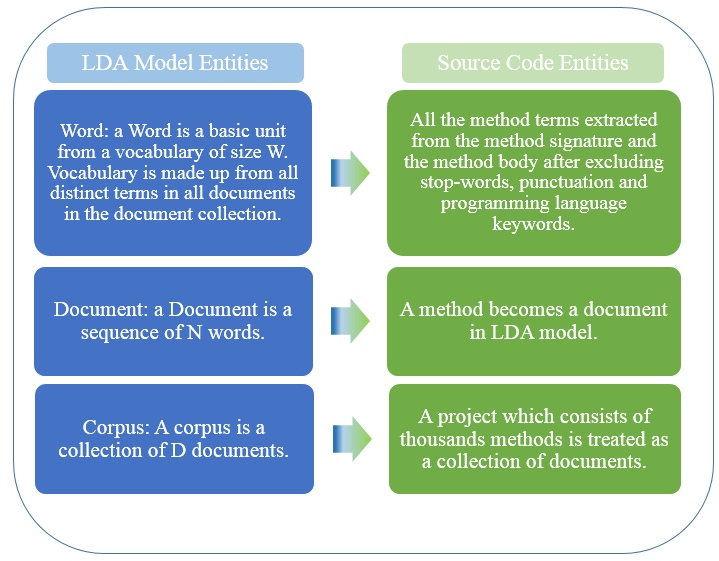}
  \caption{Mapping LDA to source code.}
  \label{fig:mapping}
\end{figure}

LDA ~\cite{blei2003latent} is a generative statistical topic modeling technique used to extract and define latent topics present in a set of documents and classify documents as a mixture of a finite set of topics by association between each meaningful word in the document and one of these topics. The LDA technique has advantages in modularity and extensibility compared to other information retrieval approaches like \emph{LSI} (\emph{Latent Semantic Indexing})~\cite{hotmann:lsi}, and this is the fact that led us to employ it for code summarization. Figure~\ref{fig:mapping} describes three parameters which must be identified to apply LDA model.

To implement the LDA component of CrowdSummarizer, we used the \emph{Mallet}'s\footnote{\underline{MA}chine \underline{L}earning for \underline{L}anguag\underline{E} \underline{T}oolkit, \url{http://mallet.cs.umass.edu/}} Eclipse plug-in.
\subsubsection{Choosing the Number of Topics\label{pat:number-of-topics}}
One of the most important and challenging parameters for constructing the LDA model is identifying the total number of latent topics which is called T. However, choosing the number of topics in the LDA model is as difficult as choosing the dimensionality reduction parameter in LSI. In the both cases the size of corpus and the size of each document are significant factors. Although, the automated algorithms are working on estimating the appropriate value, but the most common solution to this problem is varying the number of topics to get the best classification from the domain experts' viewpoints{maskeri2008mining}.
Depends on the number of documents in the document collection, researches indicated a value for T in the range of 50 to 300. Here are examples of the most commonly used and successful values for different sizes of the document collection. T=100 for D \textless= 20,000 and T=300 for 30,000 \textless D \textless 40,000. Studies for identifying the number of topics for collections greater than 50,000 is still ongoing~\cite{linstead2007mining}.
From the 32 open-source Java programs (Table \ref{tab:number-of-methods}), we analysed that the average number of methods in the Java programs is about 10,000 (min = 318 and max =39,747). Thus, the T value for a Java programs is often less than 100 and it is a function of the number of methods in that program.

\begin{table}[h]
	\centering
	\small
    \caption{The number of methods in different Java applications.}
    \label{tab:number-of-methods}
\newcolumntype{s}{>{\columncolor[HTML]{AAACED}} c}
\definecolor{green}{RGB}{112, 173, 71}
\resizebox{\textwidth}{!}
{\begin{tabular}{ |c|s||c|s||c|s| }
\hline
Program & No. Methods & Program & No. Methods & Program & No. Methods  \\
\hline\hline
JTopas & 613 & Megamek	& 9300	& HsqlDB & 5150\\
\hline
Jajuk & 5921 & SweetHome3D & 4083 & JBidWatcher & 1877\\
\hline
JEdit & 7161 & Freecol & 5971 & PlanetaMessenger & 1142\\
\hline
JHotdraw & 5263 & GanttProject & 4956 & Freemind & 6110\\
\hline
Hibernate & 12793 & Jabref & 5368 & JavaHMO & 1737\\
\hline
JFtp & 2379 & MegaMek & 9256 & Vuze & 36372\\
\hline
aTunes & 1852 & Art of Illusion  & 7727 & ArgoUML & 10341\\
\hline
Ant & 9146 & JasperReports & 12349 & JfreeChart & 8230\\
\hline
Cactus & 926 & Tomcat & 11394 & NanoXML & 318\\
\hline
Liferay & 39747 & OpenOffice & 20374 & iText & 34141\\
\hline
Tiger Envelopes & 3005 & Azureus & 28255 &\cellcolor{green} Average &\cellcolor{green} 9,892\\
\hline
\end{tabular}}
\end{table}
%%%%%%%%%%%%%%%%%%%%%%%%%%%%%%%%%%%%%%%%%%%
%%%
%%%  Generating Summaries for the Methods via Crowdsourcing
%%%
%%%%%%%%%%%%%%%%%%%%%%%%%%%%%%%%%%%%%%%%%%%

\subsection{Collecting Summaries for the Methods via Crowdsourcing}\label{sec:crowdsourcing}
The term crowdsourcing is a combination of the terms ``crowd" and ``outsourcing". The crowdsourcing approach is a novel problem-solving technique that aims to outsource different tasks to a crowd of people through the open call (e.g., via the Internet) instead of traditional suppliers~\cite{howe-rise}. Crowdsourcing has been applied in a wide range of researches, such as social networks~\cite{forlines}, mobile~\cite{eagle}, health~\cite{prpic}, and etc. These tasks often cannot be solved or are difficult to be processed by computers, but are simple enough to be performed by workers.

In recent years, crowdsourcing has attracted significant attentions to support a wide range of software engineering activities to like requirements engineering, design, coding, software testing and maintenance~\cite{mao-survey}. Examples of the platforms that are specifically used in software engineering field are TopCoder, uTest, AppStori, Bountify which are working in software development, software testing mobile application development, respectively~\cite{mao-survey}.

As mentioned earlier, code summarization helps developers to have a general understanding of a source code in a shorter period of time, particularly when it is a complex one and includes thousands of lines of code. Nonetheless, source codes often suffer from the lack of proper comments, documentation, and summaries, since manually creating them requires a lot of effort, and developers are often hesitant to perform it. Consequently, automated code summarization approaches, such as~\cite{rodeghero, sridhara-automatic-generation:PhD:2012, McBurney:TSE:2016}, have been introduced in literature. However, as will be discussed in \secref{relatedwork}, they pose some limitations like the limited length of the methods they can summarize, or they can only summarize some special aspects of a program (e.g., the contexts or the roles of methods). Hence, to mitigate these problems, we propose to use the power of the crowd by adapting the crowdsourcing technique for code summarization. More specifically, we developed the CrowdSummarizer platform with which developers can decompose a Java program into its methods, and submit those methods to a website. Next, a crowd of developers would write summaries for those methods, and the results are then presented to the user. To save user's time and to give them some initial results, as will be discussed later in \secref{automatic_summary}, CrowdSummarizer also automatically generates summaries for those methods with the help of the knowledge it has learned so far.

In order to develop a crowdsourcing platform we must identified the three main components of a crowdsourcing system defined by~\cite{estelles2012towards}:
\begin{itemize}
\item The Crowd (Workers): any developer can participate and supposed to write summaries for the methods and evaluating other summaries and in return he will get points, badges and prizes.
\item The Requester: a person who post a code to the platform and wants other to write summaries for the code.
\item Process: Type: delegation a task to the crowd of users. Call: open call. Medium: the internet.
\end{itemize}

\begin{figure}[h]
  \centering
  \includegraphics[width=400px]{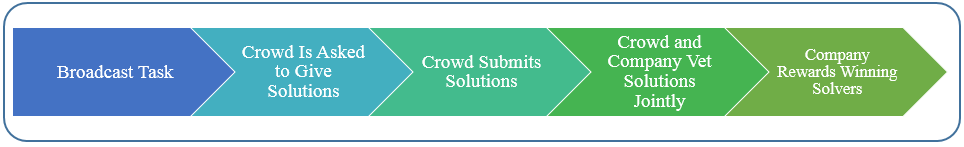}
  \caption{A typical Crowdsourcing process}
  \label{fig:crowd-process}
\end{figure}

Figure \ref{fig:crowd-process} shows a typical crowedsourcing process. After broadcasting a method from a requester, workers(developers) are asked to write summaries for that method and submit their answers on the platform. Examining solutions carefully to make decision to accept the answers is one of the important steps in this process. As discussed earlier, although gamification makes users to write high quality summaries, but to discourage users from writing unrelated comments, we added an evaluation part to the platform which works by giving positive or negative feedbacks to summaries (the way that StackOverflow does voting answers). We show methods with correspondent summaries(to prevent personal biases we don't display authors of the summaries) to developers and ask them if summaries are related to the method or not. Using this mechanism, unrelated summaries will be removed automatically by identifying summaries with negative feedbacks. We consider a point mechanism to encourage users to evaluate summaries.

With respect to the above discussion, The most fundamental issue and the key requirement in our work is to get people to take part in crowdsourcing and have a sheer number of users to write a summary for a piece of the code.  In some cases, the workers join to a platform because of their personal interest, altruism or maybe just because of they believe in you. However, how to be sure about the quality of the work outcome? It has been a weakness of crowdsourcing methodology due to its open participation model~\cite{morschheuser2016gamification}. To mitigate this weakness, motivation of the workers is of a great importance. Although, task characteristics and granularity have also impact on worker's motivation, but there are two kinds of motivations: intrinsic and extrinsic~\cite{feyisetan2015improving}. Feeling competent and creative, having fun, volunteering projects and achieving social recognition are examples of intrinsic motivation, while financial payoffs, such as gift cards is extrinsic motivation. Researches indicated that the combination of these two factors will increase the participation. One of the great drivers for increasing intrinsic motivation and engaging users in a crowdsourced task is \emph{gamification}~\cite{morschheuser2016gamification} which is discussed in the following.
\subsubsection{Gamification}\label{sec:gamification}
Gamification is a neologism from the word ``game" and embraces the concept of using the game design thinking and mechanisms in non-game contexts in order to make them more fun and engaging~\cite{zichermann2011gamification}. Gamification makes people to enjoy their pastime while performing the crowdsourced tasks. To this end, as mentioned above, to motivate developers to engage in writing summaries for the methods submitted to the CrowdSummarizer's website, we designed and implemented a web-based game as follows (see~\cite{supportingmaterials} for a demonstration of this game).

\begin{figure}[h]
  \centering
  \includegraphics[width=\textwidth]{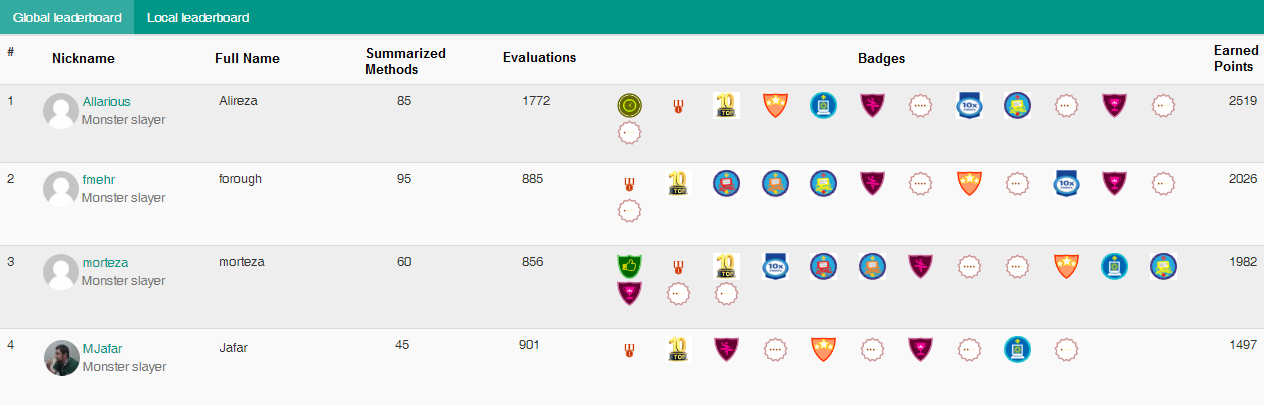}
  \caption{An screenshot from the CrowdSummarizer's global leaderboard.}
  \label{fig:leaderboard}
\end{figure}

\begin{figure}[h]
  \centering
  \includegraphics[width=\textwidth]{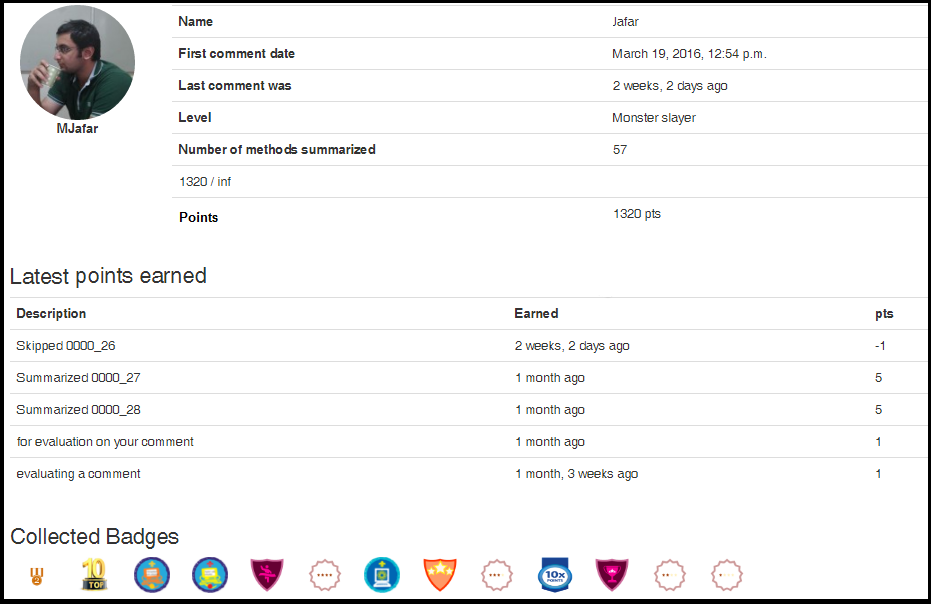}
  \caption{An screenshot from the CrowdSummarizer's user's profile.}
  \label{fig:profile}
\end{figure}

Any developers who registers in the CrowdSummarizer platform can be a player in the game, can earn points, and move up to the top players list. This game includes eight levels which starts with summarizing simpler methods, and then continues with summarizing harder ones in the next levels. Evaluating submitted answers carefully to decide whether or not an answer should be accepted is an important step in any crowdsourcing process. Therefore, we added an evaluation step to the game which randomly shows a number of methods together with their submitted summaries to players, and allows them to \emph{upvote} or \emph{downvote} other players' summaries. To prevent personal biases, we hide the names of the authors of those summaries. In this way, irrelevant summaries will be removed automatically by identifying summaries with negative votes. To motivate further the developers to engage in the game, players must reach at least the level \#4 %% Maybe someone wants to pay some money. But, he himself is not interested to play!!!
of the game to be allowed to push their own code for summarization to the crowd. Moreover, we used the following different types of gamification elements in our platform to make the game more fun and motivating:
\begin{itemize}
\item \emph{Points}:
It is the core element used in almost any gamified system as rewarding mechanism for quantitative number of fulfilled enumerable tasks that informs users about more valuable tasks. With respect to its difficulty, each method has a point that if a player write a summary for it, she would get it. Moreover, the first three players who summarize a method would receive double points. We randomly select several methods as starred methods which have extra points. Evaluating other players' summaries, and also having summaries with high upvotes can also increase the total points. Obviously, having summaries with downvotes can decrease the total points. To prevent players from worthless evaluations of others' summaries, we use a trap mechanism and a fake account to identify cheaters.

\item \emph{Leaderboards}: Researches on gamification indicated that ranking systems are very effective in motivating players~\cite{zichermann2011gamification}.
 Thus, we employed two different leaderboards in our game: a \emph{global} leaderboard and a \emph{local} leaderboard. The former shows the competition amongst all the players (\figref{leaderboard}), while the latter just displays the players who are in that player's programming experience level. We also present encouraging statements to players like "Hurry up, writing 2 summaries will shift you up to the second place in the global leaderboard.", "You missed your place in the local leaderboard. It is the time to get it back.", "Good job!", or "keep on!".
\item \emph{Badges}: Players get different tagged images as badges based on their performance in the game ((\figref{leaderboard})). For example, a player with the highest upvotes will get the ``Good summarizer" badge. Example of other badges are "Quick summarizer", "Newbie", "Adventure", "Explorer", "Superstar", and etc.

\item \emph{Levels}: Generally, levels indicate where players stand in a game~\cite{zichermann2011gamification}. Users will pass 8 levels according to their points in a non-linear form. We use progress bar which shows the percentage of completeness of the level as well as levels to show player's own advancement. To increase the attractiveness of the game, we considered a title for each level. For example, ``Starting to see the light", ``Middle of the way", and ``Monster slayer" for levels \#1, \#4, and \#8 respectively.

\item \emph{Mystery Boxes}: We show the players a mystery box with a random gift (e.g., point, badge, etc.) in levels \#2, \#5, and \#7 to motivate them to play further.

\item \emph{Avatars}: Representing players with an avatar can increase their engagements~\cite{zichermann2011gamification}. Thus, if a player has set an avatar in \emph{Gravatar}\footnote{A Globally Recognized Avatar, \url{https://en.gravatar.com/}}, we would use it in our game (Figure \figref{profile} shows the user's profile.).
\end{itemize}

Results indicated that points and leaderboards are the two most common and successful affordances which employed in crowdsourcing contexts with homogeneous contributions~\cite{morschheuser2016gamification}. Despite all these elements and the positive results which gamification has in the crowdsourcing process, but there is still a risk. Using too much of these elements may makes workers just concentrate on receiving points and getting higher ranks in the leaderboards rather than performing the summarization task. However, we are trying to stop getting displaced scores by using evaluation and trap mechanisms.

\subsubsection{Getting Summaries from the CrowdSummarizer Platform \label{sec:getting_summary}}
We have described so far in this section that users of the CrowdSummarizer platform submit their desired methods of a Java program as code summarization tasks to our web-based game, and a set of players submit their own summaries for those methods, and the summaries are then evaluated by another set of players. Verifying tasks done by the crowd is one of the challenges in any crowdsourcing platforms that we addressed it in our approach by the crowd itself. Afterwards, users can choose different kinds of outputs from the collected summaries. Currently, CrowdSummarizer supports the following three kinds of outputs:

\begin{itemize}
\item \emph{Summary with the highest upvotes:} With respect to the number of upvotes and downvotes of submitted summaries for a given method, a summary is chosen.

\item \emph{Summary with the highest similarity with the given method:} A summary that has the most common keywords with the given method is selected (see \eqaref{commonwords} in \secref{compweights}).

\item \emph{Merged summaries:} This extracts the common concepts amongst the submitted summaries. Users also have the option to access each submitted summary if needed.
\end{itemize}

\subsection{Automatically Generating Summaries\label{sec:automatic_summary}}
In the previous section, we used the CrowdSummarizer platform in an active form to get the summaries from a number of players. However, in this section, we look into this platform as a passive component. More specifically, CrowdSummarizer continuously learns a set of weights and sentence templates from the set of methods and their corresponding summaries which have been collected from the crowd so far, and uses them to automatically generate summaries for new methods. This helps developers to get some initial results from the platform, and to not necessarily wait for the crowd to write summaries for their methods. In the following, we describe the steps that are followed to generate automated summaries. McBurney et~al.~\cite{mcburney-empirical} pointed out that the source code and the readers' summaries are more similar than the source code and the authors' summaries. Authors use more development details and low level implementation informations. Meanwhile, readers are those who want to understand the code; therefore, they must be able to deduce the concepts from the low level details. So the ultimate goal of source code summarization approaches must be generating summaries that are more similar to summaries that a reader would write.
\subsubsection{Calculating the Weights of Method Parts for Being in Summaries}\label{sec:compweights}
The aim of this step is to assign weights to different parts of a method (e.g., method's name and parameters) with respect to the frequency of their use in the summaries collected so far by the CrowdSummarizer platform. These weights are then used to automatically generate a summary for a given method.

\begin{figure}[h]
  \centering
  \includegraphics[width=400px]{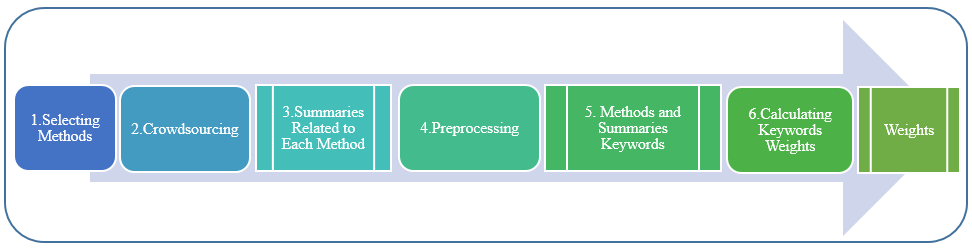}
  \caption{Computing word weights using Crowdsourcing results.}
  \label{fig:weight process}
\end{figure}

Overall process of computing keyword weights is shown in Figure \ref{fig:weight process}. The output of this process is a set of keyword weights which will be used in automatically generating summary for a method.

Similar to other approaches which are based on linguistic information, the quality of our results are significantly affected by meaningless identifiers. Thus, our approach produces better results when meaningful identifers are used, and naming conventions and style guidelines are followed. Additionally, like other natural language processing systems, we apply several general preprocessing steps to refine the data set and to prepare it for the next analysis tasks:

\begin{itemize}
\item \emph{Tokenization:} This step tokenizes a sentence into its constituent words.

\item \emph{Splitting:} As readers use method keywords in their summaries and these keywords usually follow certain coding conventions, we had to split these words. We implemented the splitter introduced in~\cite{Lawrie} which is based on common coding styles, such as camel cases, punctuation, numbers, underscores, and so on. For example, the term ``print\underline{\space}mp3FileContent" would be splitted into the ``print", ``mp", ``3", ``File", and ``Content" terms. However, for the Java programs properly following the coding standards, performance of a simple CamelCase is adequate~\cite{guerrouj2012tris}.

\item \emph{Expanding abbreviations:} Abbreviations (e.g., ``pntr" and ``ptr") must be expanded and turned into the complete words (e.g., ``pointer") to increase consistency. We use the \emph{AMAP} approach~\cite{hill2008amap} to automatically identify and expand abbreviations.

\item \emph{Transforming to lowercases:} We make all the words case-insensitive to increase consistency.

\item \emph{Correcting the spelling of words:} We noticed in our evaluations of CrowdSummarizer that spelling errors are quite common in collected summaries. Some users correct spelling errors using browser's spell checkers(e.g. the way that Google does spelling correction~\cite{whitelaw2009using}), and some don't. Hence, we use the spelling corrector provided by P. Norvig\footnote{\url{http://norvig.com/spell-correct.html}} to fix the errors. The core of all the spell checkers is a big text file which consists of about a million words from several public domains. Using probability theory the algorithm chooses the most likely spelling correction for a word.

\item \emph{Removing stopwords:} As an important step in preprocessing the data set, we removed common English language stop words like ``the", ``but", ``a", ``or", and so on.

\item \emph{Stemming:} Stemming is the process of reducing derived words into their common word stem or root. An example of stemming is transforming the terms ``cat", ``cats", and ``kittens" to the term ``cat". There are different stemming algorithms, such as Snowball's Stemmer, Lancaster's Stemmer, Porter's Stemmer and WordNet's Lemmatizer, that reduce words to their morphological root. We use \emph{WordNet Lemmatizer}\footnote{\url{http://www.nltk.org/api/nltk.stem.html}} for this purpose.
\end{itemize}

We carried out the above steps to extract the keywords of each method $m$, and each of its submitted summaries $s$, called $MethodKeywords(m)$ and $SummaryKeywords(s)$, respectively. Thus, for a method $m$, the common keywords between that method and its corresponding summary $s$ can be computed as follows:

\begin{equation}
CommonWords(s,m) = {SummaryKeywords(s) \cap (MethodKeywords(m) \cup Syn(MethodKeywords(m)))}
\label{eq:commonwords}
\end{equation}

$Syn(...)$ is the set of all the synonyms of the words in $MethodKeywords(m)$. We used the \emph{WordNet}\footnote{\url{https://wordnet.princeton.edu}} lexical database to find the synonyms of each word. For example, suppose the words ``calculate" in the name, ``bike" in the parameters list, and ``copy" in the body of a method. While summarizing that method, some developers might have used exactly the same words, while others might have used their synonyms like ``compute", ``bicycle", and ``duplicate", respectively. Consequently, these pairs of words (i.e., (``calculate", ``compute"), (``bike", ``bicycle"), and (``copy", ``duplicate")) must be considered the same when we calculate the degrees of importance of different parts of a method to be present in its corresponding summary. Hence, we take into account the synonyms of words to increase the precision of results.

To compute the weight of each word in the set of $CommonWords$, we first calculate the weight of that word in that method and each of its summaries in the following way:

\begin{equation}
Weight\_in\_method(w, m) = \frac{\textit{The number of occurrences of the word w in $MethodKeywords(m)$}}{\textit{The number of words in $MethodKeywords(m)$}}
\label{eq:eq_wieght_method}
\end{equation}

\begin{equation}
 Weight\_in\_summary(w, s) = \frac{\textit{The number of occurrences of the word w in $SummaryKeywords(s)$}}{\textit{The number of words in $SummaryKeywords(s)$}}
\label{eq:eq_wieght_summary}
\end{equation}

Then, for each word in the $CommonWords$, we compute a normalized weight, which means the rate of using the method words in a summary, as follows:

\begin{equation}
NormalizedWeight(w)= \frac{Weight\_in\_summary(w, s)}{Weight\_in\_method(w, m)}
\end{equation}

Next, $weight(w)$ is the average of normalized weights for the keyword $w$ amongst all the summaries submitted for a method. After this, we map each keyword to various parts of the methods to see which parts are more frequently used in the summaries.

In our evaluations of CrowdSummarizer with 149 developers of various levels of experience (see \secref{eval}), we noticed that various properties of methods impact on the weights of different code parts. Thus, we categorised the methods based on the following metrics:

\begin{itemize}
\item \textit{Length of lines}: Small(3 \textless= LOC \textless= 20), medium (20 \textless= LOC \textless= 70), large (LOC \textgreater= 70)
\item \textit{Length of parameter lists}: None, 1 \textless= no. parameters \textless= 3, 3 \textless no. parameters.
\item \textit{Type of return value}: Vector, Boolean, numbers (Integer, Float and etc.), string (char or an array of chars), object and etc.
\item \textit{Methods with the highest fan-in}: Methods which are called the most.
\item \textit{Methods with the highest fan-out}: Methods which call too many methods.
\item \textit{The type of work that it does }(\textit{Method stereotype}\cite{Dragan}): Interacting, command and collaborator methods.
\item \textit{Static methods}.
\end{itemize}

Additionally, we noticed that developers mainly consider the following parts of a method, in the following order of importance, when they write their summaries:

\begin{enumerate}
\item Method's Name and Return Type,
\item Parameters,
\item Ending Units (e.g., \emph{return} statements, \emph{printout} statements, etc.),
\item Method Invocations,
\item Branches (i.e., \emph{if} and \emph{switch} statements),
\item Loops (i.e., \emph{for}, \emph{while}, and \emph{do-while} statements),
\item Assignments,
\item Local Variables, and
\item Error Handlings (e.g., \emph{try-catch}, \emph{exception}, \emph{throw}, etc.).
\end{enumerate}

Keywords in the method are tagged with the above ordered list (i.e. if a keywords belongs to both \textit{Method Name} and \textit{Local Variable}, we count it in \textit{Method Name} area), using Abstract Syntax Tree\footnote{Java Parser, \url{https://github.com/javaparser/javaparser/wiki/Manual}}(AST). Finally, we have an average weight for each of these items according to different properties of methods which we discussed earlier. \tabref{weights} shows the weights of the three categories of methods.

\begin{table}[h]
	\centering
	\small
    \caption{The term weights for different code areas for different method categories.}
    \label{tab:weights}

\resizebox{\textwidth}{!}
{\begin{tabular}{ |c|c|c|c|c|c|c|c|c|c|}
\hline
\rowcolor[HTML]{AAACED} Method Categories & Method Name & Parameter & Ending Units & Method Invocation & Branches & Loops & Assignments & Local Variable & Error Handling \\
\hline
collaborator & 5.8	& 3.9	& 4.25 & 3.09 &  1.78	& 1.49	& 1.87	& 2.16	& 0.5	\\
\hline
highest fan-in	& 5.17	& 2.37	& 3.26 & 2.44 &  2.52 & 0.72 & 1.31	& 1.66 & 1.69\\
\hline
highest fan-out	& 4.85	& 3.27	& 4.62 & 4.63 &  2.76 & 1.2 & 2.26 & 2.28	& 0.65\\
\hline
\end{tabular}}
\end{table}
If a method belongs to different categories we calculate average weight for its different code areas.
\subsubsection{Selecting the Most Important Keywords}
So far, we have keyword weights for the different portions of the method for different kinds of the methods. The goal of this step is to select the most important keywords for describing a method. In information retrieval, \emph{tf idf} (\emph{term frequency inverse document frequency})~\cite{salton-automatic}, is a numerical statistic that reflects how important a word is to a document in a corpus. The first part(i.e. \emph{tf}) gives us importance of a word in a document by counting the number of times that the word occurs in the document. The latter is about the words which appear in the large number of documents, so it seems that these are not important for this specific document~\cite{binkley2010information}. In the other words, the goal is to give the high weights to the rare words and the low weights to the frequent words. The \emph{tf idf} factor is widely used in natural language text processing and also in text summarization~\cite{haiduc-use}. However, the problem with using it in code summarization is that it treats the source code as a plain text, and hence, the words in different parts of a method are treated equally. For example, the words in the method signature are treated the same as the words in a \emph{while} loop. However, we discussed in the previous section that developers distinguish between various parts of a method. To address this issue, we also consider the weight factor computed in the previous step. More specifically, the importance of a term $t$, in a method $m$, in the program $p$ can be computed using the semicode displayed in Listing \ref{lst:importance}.
\begin{lstlisting}[language=C++,label={lst:importance}, caption=Calculating the Importance of a Term in a Method]
CalcImportanc(Term t, Method m, Program p)
{
  tf = The number of occurrences of t in m;
  N = The number of methods in p;
  df= The number of methods in p which t occurs in;
  idf = log (N/df);
  tf-idf= (1+ log tf)*idf;
  weight = getWeightFromWeightsDB(t,m);
  importance = tf-idf * weight;
  return importance
}
int getWeightFromWeightsDB(Term t, Method m)
{
  methodCategory = identifyMethodCategory(m):// which category is this method belongs to according to method properties.
  termArea = identifyTermAreaInMethod(t);// which 9 area is this term belongs to
  return (weight(methodCategory,termArea));
}
\end{lstlisting}

In short, based on the observation that we had on the length of the crowd summaries(see Section \ref{discussion}), the number of keywords selected for the summary of a method is related to the complexity of that method~\cite{Scalabrino}.
\subsubsection{Using Templates to Generate Natural Language Summaries}
\begin{figure}[h]
  \centering
  \includegraphics[width=\textwidth]{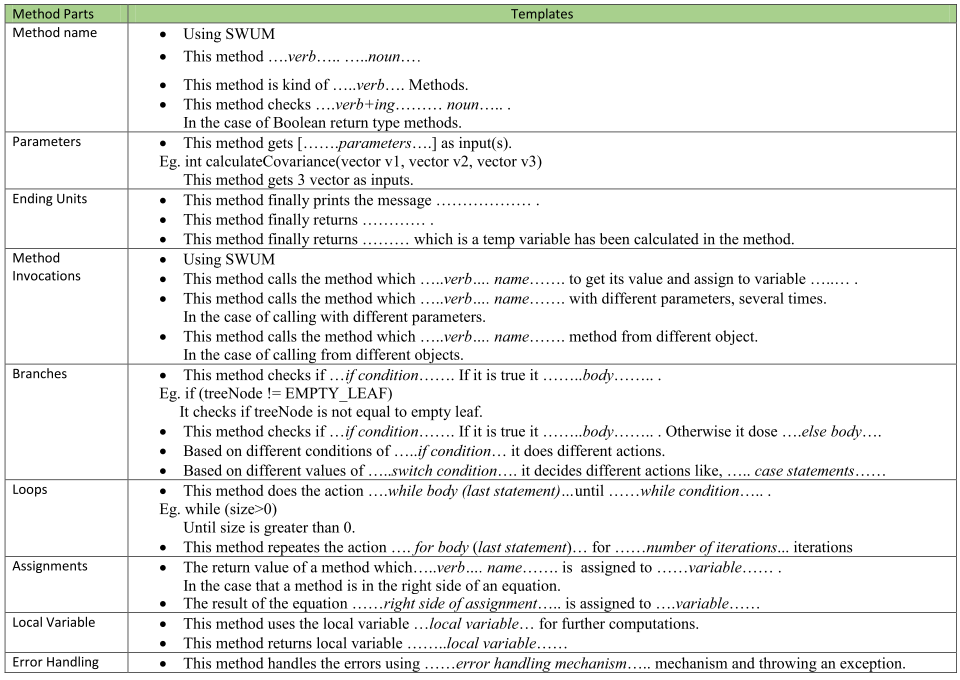}
  \caption{Templates used in generating natural language summaries in CrowdSummarizer.}\label{fig:templates}
\end{figure}

The last step of our approach is to automatically generate the summaries of methods in the form of natural language sentences. For this purpose, we use a number of templates that were extracted from the summaries written by the crowd in the CrowdSummarizer platform for different kinds of methods. For instance, it was observed in our evaluations that developers in 90\% of the time use the terms ``till" or ``until" instead of the Java keyword \emph{while} in their summaries. As another example, the methods that return a value of type \emph{boolean}, in 86\% of cases has the term ``check" in their summaries. If the method has a list of parameters, for example \textit{v1, v2} and \textit{v3}, users write just the last index of variable in their summaries (i.e. three parameters). For conditions, \textit{while, if, case, for} and etc., instead of operators, like ==, !=, \textless and \textgreater, the words \textit{equal}, \textit{not} \textit{equal}, \textit{smaller}(lower, less than) and \textit{greater}(higher, bigger) are used, respectively. There is a great deal of this information which we used to generate templates for each portions of the method. The further details about the templates are shown in \figref{templates} .

%%%%%%%%%%%%%%%%%%%%%%%%%%%%%%%%%%%%%%%%%%%
%%%
%%%             Implementation
%%%
%%%%%%%%%%%%%%%%%%%%%%%%%%%%%%%%%%%%%%%%%%%
\section{CrowdSummarizer Implementation \label{sec:impl}}
The CrowdSummarizer approach introduced in this article is fully automated and is implemented as an Eclipse plugin as well as a web-based game which is implemented in \emph{Python}. The CrowdSummarizer's Eclipse plugin takes a Java program as input and automatically outputs the hierarchical structure of the latent topics in that program (see \secref{topic-model}) as well as the natural language summaries generated automatically for the methods of that program (see \secref{automatic_summary}). Furthermore, upon user's request, it submits those methods to CrowdSummarizer's web-based game for the summarization by the crowd (see \secref{crowdsourcing}). A demonstration of our implementations of CrowdSummarizer is available online~\cite{supportingmaterials}.
%%%%%%%%%%%%%%%%%%%%%%%%%%%%%%%%%%%%%%%%%%%
%%%
%%%             Example
%%%
%%%%%%%%%%%%%%%%%%%%%%%%%%%%%%%%%%%%%%%%%%%
\section{Example\label{exp}}
\begin{figure}[h]
  \centering
  \includegraphics[width=\textwidth]{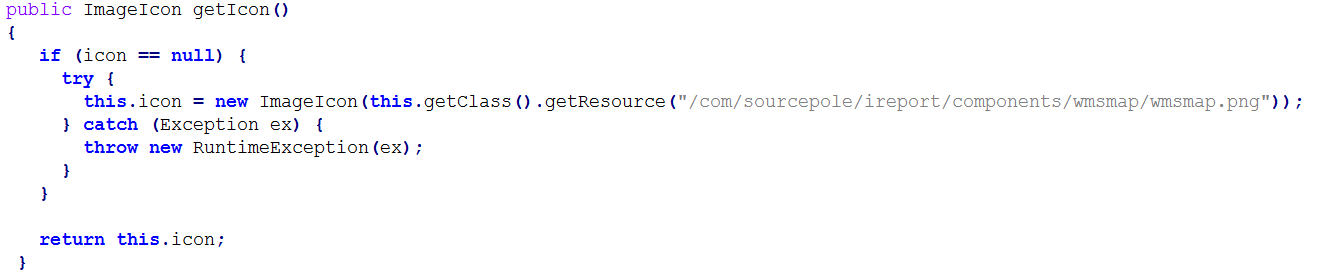}
  \caption{A method example for generating summary by CrowdSummarizer.}\label{fig:example-method}
\end{figure}

\begin{figure}[h]
  \centering
  \includegraphics[width=0.55\textwidth]{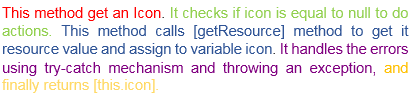}
  \caption{An example of a summary automatically generated by CrowdSummarizer for a method shown in \figref{example-method}}\label{fig:example-summary}
\end{figure}

In this section, we explore an example of automatically generating a summary for a specific method. Consider the method $getIcon()$ from the application JasperReports (see \figref{example-method}).
 At first we perform preprocessing steps discussed in \secref{automatic_summary} to refine the method keywords. Now for each of the refined keywords we must compute its importance factor(see Listing \ref{lst:importance}). We run JavaPaser on this method to extract each area which each keyword belong to. Additionally, we consider the properties of the method(see \secref{automatic_summary}) to identify the method category. This method has no attribute list and 7 code lines. Using the program call graph we understand that this method is kind of \emph{collaborator} method which called so many times by the other methods. Thus we use the average weights for the method keywords using these two kinds of data. The next step is to select the top-n keywords. As we discussed in \secref{automatic_summary} and Section \ref{length} the lengths of the summaries are affected by the method complexity. We compute its complexity by considering method lines, conditional statements and etc. \cite{Scalabrino} which results the value 5 for this method. Therefore, we select the following top 5 keywords to generate the summary:
\begin{enumerate}
\item \emph{get Icon} : method name
\item \emph{icon == null} : branches
\item \emph{get Resource} : method invocation
\item \emph{try catch} : exception mechanism
\item \emph{this . icon} : ending unit
\end{enumerate}
We use the templates for each code area and put the sentences together to generate a paragraph for describing the method. The fully combined
summary is shown in \figref{example-summary}.

%%%%%%%%%%%%%%%%%%%%%%%%%%%%%%%%%%%%%%%%%%%
%%%
%%%             Evaluation
%%%
%%%%%%%%%%%%%%%%%%%%%%%%%%%%%%%%%%%%%%%%%%%
\section{Evaluations\label{sec:eval}}
The CrowdSummarizer platform was evaluated by two user study experiments. The first one evaluates the applicability of the crowdsourcing approach for code summarization (\secref{eval:CSApplicability}), while the second one evaluates the quality of automatically generated summaries (\secref{eval:CSResultsQuality}).

%%%%%%%%%%%%%%%%%%%%%%%%%%%%%%%%%%%%%%%%%%%%%%%%%%%%%
%%%
%%%  Evaluation of the Applicability of CrowdSummarizer
%%%
%%%%%%%%%%%%%%%%%%%%%%%%%%%%%%%%%%%%%%%%%%%%%%%%%%%%%
\subsection{Evaluating the Applicability of CrowdSummarizer}\label{sec:eval:CSApplicability}

%%%%%%%%%%%%%%%%%%% OBJECTIVES %%%
\subsubsection{Objectives}
The main research question that this empirical study intends to answer is ``whether the CrowdSummarizer platform is applicable and motivating for developers to use it in practice?"

%%%%%%%%%%%%%%%%%%% SETUP %%%
\subsubsection{Setup}
\paragraph{Methodology.}

At the end of the game, we asked participants to fill out a questionnaire which consisted of two parts. The first part included the questions about the applicability and usefulness of the game in practice, and the ways in which our proposed approach could be improved. However, the second part was about opinions about source code summarization and different factors impact on it.
\paragraph{Selecting of Methods.}

We collected about 4000 summaries for 128 different methods, with an average of 19 summaries for each method. To increase the generalizability of the results, the methods were selected from 11 different applications from various domains (e.g., text editors, multimedia, games, etc.) of different sizes (including 318 to 12793 methods). Moreover, from each application, a set of methods with various properties were chosen (e.g., number of lines of code, number of parameters, type of return values, etc.). See Table \ref{tab:lenght_method} for more details.
\paragraph{Participants.}

To answer the above research question, we asked 149 developers, with an average of 5.29 years of programming experience, to play the CrowdSummarizer's web-based game. The details are shown in Table \ref{tab:number-of-Participants-1}.

\begin{table}[h]
	\centering
	\small
    \caption{Participants in initial examination of CrowdSummarizer game.}
    \label{tab:number-of-Participants-1}

\resizebox{\textwidth}{!}
{\begin{tabular}{ |c|p{3.5cm}|p{1.6cm}|p{2.4cm}|p{2.3cm}||p{2.3cm}|p{2.3cm}|p{2.7cm}| }
\hline
\rowcolor[HTML]{AAACED} \multicolumn{8}{|c|}{Number of Participants} \\
\hline
No.Months & General Programming  Experience	& Industry Experience & C++ Experience & Java Experience & Java Self-Assessment & C++ Self-Assessment & Proficiency Level 5=Highest\\\hline
84+ months & \centering\arraybackslash 36 & \centering\arraybackslash 6 & \centering\arraybackslash 10 & \centering\arraybackslash 15 & \centering\arraybackslash 14 & \centering\arraybackslash 20 & \centering\arraybackslash 1\\
48-84 months & \centering\arraybackslash 41 & \centering\arraybackslash 12 & \centering\arraybackslash 45 & \centering\arraybackslash 41 & \centering\arraybackslash 32 & \centering\arraybackslash 33 & \centering\arraybackslash 2\\
24-48 months & \centering\arraybackslash 36 & \centering\arraybackslash 28 & \centering\arraybackslash 48 & \centering\arraybackslash 56 & \centering\arraybackslash 37 & \centering\arraybackslash 36 & \centering\arraybackslash 3\\
6-24 months & \centering\arraybackslash 34 & \centering\arraybackslash 56 & \centering\arraybackslash 29 & \centering\arraybackslash 27 & \centering\arraybackslash 27 & \centering\arraybackslash 23 & \centering\arraybackslash 4\\
0-6 months & \centering\arraybackslash 2 & \centering\arraybackslash 47 & \centering\arraybackslash 17 & \centering\arraybackslash 10 & \centering\arraybackslash 39 & \centering\arraybackslash 16 & \centering\arraybackslash 5\\
\hline
\end{tabular}}
\end{table}

\paragraph{Metrics.}

Questions are kinds of multiple choice questions which we assigned a values to the answers as higher value for stronger performance and preferred answer.

%%%%%%%%%%%%%%%%%%% RESULTS %%%
\subsubsection{Results}
\paragraph{CrowdSummarizer applicability results.}

Table \ref{tab:applicability} summarizes the feedbacks developers gave us at the end of the game about the CrowdSummarizer platform. As can be understood from these results, CrowdSummarizer is applicable and motivating for developers to use it in practice. In addition, users' answers to open-ended question also reflected their positive feeling about the CrowdSummarizer game. Here is some answers:

{\it{"First of all I want to say that gamification elements such as, badges, points and etc., had increased attractiveness of the game we were playing and are strength of the projects.}} {\it{Even with making these elements more complicated, you can attract more fans."

"The strategic plan of the site is very good and provocative. Try to improve scoring mechanism."}}

Although, we got more positives responses, but there are still rooms for improvement in some aspects of the approach. For instance, users felt that the evaluation part of the game was a little boring.

\begin{table}[h]
	\centering
	\small
    \caption{The results of evaluating the applicability of CrowdSummarizer}
    \label{tab:applicability}
\definecolor{green}{RGB}{112, 173, 71}
\newcolumntype{s}{>{\columncolor{green}} c}
%\resizebox{\textwidth}{!}
{\begin{tabular}{ |p{7cm}|c|c|c|c|c|s|}
\hline
 \rowcolor[HTML]{AAACED} & \multicolumn{6}{c|}{points}  \\
\cline{2-7}
 \cellcolor[HTML]{AAACED} Questions & 1 & 2 & 3 & 4 & 5 & Avg \\
\hline
How enjoyable is it to play?(1=least, 5=most) & 10 & 13 & 30 & 41 & 55 & 3.79\\
\hline
How easy/difficult is it to play?(1=most difficult, 5=easiest) & 9 & 16 & 49 & 52 & 23 & 3.42\\
\hline
How successful was the website in encouraging you to summarize the codes?(1=poor,5=excellent) & 7 & 12 & 19 & 42 & 69 & 4.03\\
\hline
How successful was the website in showing the importance of code summarization?(1=poor,5=excellent) & 10 & 14 & 54 & 41 & 30 & 3.44\\
\hline
How successful was the website in enhancing your code summarizatoin skill?(1=poor,5=excellent) & 10 & 13 & 42 & 49 & 35 & 3.57\\
\hline
\end{tabular}}
\end{table}
\paragraph{Participants' Opinions About Source Code Summarization.}

Questions and related answers about source code summarization from participants' viewpoints are shown in Table \ref{tab:questionnaire_results_opinion_1} and Table \ref{tab:questionnaire_results_opiopn_2}.

\begin{table}[h]
	\centering
	\small
    \caption{Questionnaire results about users' opinions about source code summarizaion}
    \label{tab:questionnaire_results_opinion_1}
\definecolor{green}{RGB}{112, 173, 71}
\newcolumntype{s}{>{\columncolor{green}} c}
%\resizebox{\textwidth}{!}
{\begin{tabular}{ |p{10cm}|s|}
\hline
\cellcolor[HTML]{AAACED} Questions & Average Point\\
\hline
How important do you think it is to summarize a code?(1=not important,5=very important)  & 3.85\\
\hline
To what degree should contextual information about a method such as its callers be included in summaries of the code?(1=not important, 5=highly included) & 3.26\\
\hline
To what extent can the file structure of the source code (i.e quality of the code) affect the quality of summarization?(1=not important,5=highly affective) & 3.46\\
\hline
How similar should the text in summaries be to the text and keywords in the source code like method names? (1=not similar,5=very similar) & 3.7\\
\hline
\end{tabular}}

\end{table}
\begin{table}[h]
	\centering
	\small
    \caption{Questionnaire results about users' opinions about source code summarizaion(\textit{Continued})}
    \label{tab:questionnaire_results_opiopn_2}
\resizebox{\textwidth}{!}
{\begin{tabular}{ |p{5cm}|p{9cm}| }
\hline
\rowcolor[HTML]{AAACED} Questions &  Answers \\
\hline
The summary contains information that helps me understand how to use the method. &  \begin{minipage}{.2\textwidth}
      \includegraphics[width=70mm, height=30mm]{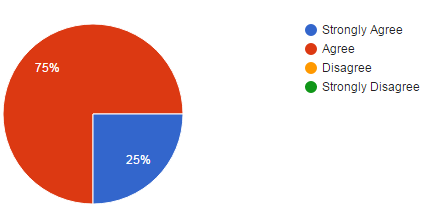}
    \end{minipage}\\
\hline
The summary contains information that helps me understand why the method exists in the program. &  \begin{minipage}{.2\textwidth}
      \includegraphics[width=70mm, height=30mm]{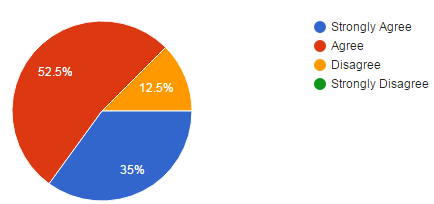}
    \end{minipage}\\
\hline
The summary contains information that helps me understand what the method does. &  \begin{minipage}{.2\textwidth}
      \includegraphics[width=70mm, height=30mm]{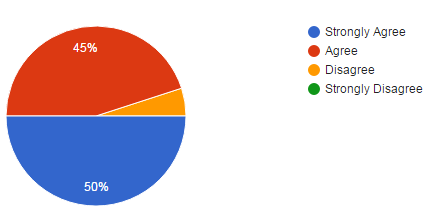}
    \end{minipage}\\
\hline
\end{tabular}}
\end{table}

%%%%%%%%%%%%%%%%%%%%%%%%%%%%%%%%%%%%%%%%%%%%%%%%%%%%%
%%%
%%%  Evaluation of the Quality of Generated Summaries
%%%
%%%%%%%%%%%%%%%%%%%%%%%%%%%%%%%%%%%%%%%%%%%%%%%%%%%%%
\subsection{Evaluating the Quality of Automatically Generated Summaries}\label{sec:eval:CSResultsQuality}

%%%%%%%%%%%%%%%%%%% OBJECTIVES %%%
\subsubsection{Objectives}
This empirical study is conducted to answer the following research questions about the automatically generated summaries:

\begin{itemize}
\item What is the quality of CrowdSummarizer's generated summaries in terms of \emph{accuracy}, \emph{comprehensibility}, and \emph{experts' satisfaction}?

\item Are the results of CrowdSummarizer better than existing code summarization approaches?
\end{itemize}

In the first research question, \emph{accuracy} means how accurate the latent topics are extracted from the program and how accurate the methods are tagged with those topics; \emph{comprehensibility} indicates how understandable the generated summaries are for developers; and \emph{experts' satisfaction} measures to what extent the developers are satisfied with the automatically generated summaries.

%%%%%%%%%%%%%%%%%%% SETUP %%%
\subsubsection{Setup}
\paragraph{Methodology.}

Our evaluation process to answer the two research questions has three phase:
\begin{enumerate}
\item To measure the accuracy of results, we gave the participants 3 Java programs from the first experiment together with 10 topmost topics and 10 methods tagged with those topics from each program . We asked them to go through the program and then evaluate the accuracy of the LDA output.
\item Afterwards, to measure the comprehensibility and satisfiability of results, we gave our experts 12 randomly selected methods from the Java programs used in the first study to 1) read the methods, and 2) write summary in their own words for each one, then, we displayed the summaries generated automatically by our approach for each method and 3) asked them to evaluate generated summary by answering the questions about comprehensibility and satisfiability of the results.
\item To answer the second research question, we compared the results of our approach with the results of the \emph{eye-tracking} approach~\cite{rodeghero} which is the closest work to ours. As we mentioned in previous step, we asked our recruited experts to write their own summaries for 12 randomly selected methods from our Java programs. Next, we compared the keywords extracted by our approach and by the eye-tracking approach to the keywords chosen by the experts in terms of \emph{precision}, \emph{recall}, \emph{F-score}, and \emph{overall accuracy}.
\end{enumerate}
\paragraph{Selecting of Programs and Methods.}

We selected (randomly) 78 methods from the same Java applications that we used in the first study. Users were assigned to see the 12 methods of the 78 methods available. We also randomly selected the 3 of these 11 applications for each participant to evaluate LDA output.
\paragraph{Participants.}

we recruited 14 expert developers who were different from those who participated in the first study (\secref{eval:CSApplicability}).
All of them are graduate students from the Computer Engineering Department at Sharif University of Technology and Isfahan University of Technology with an average of 5.5 years experience in Java programming, 8 years in general programming and  industry experience, ranging between 1 and 7 years (see Table \ref{tab:number-of-participants-2}).
\begin{table}[h]
	\centering
	\small
    \caption{Participants in examination of the quality of automatically generated summaries.}
    \label{tab:number-of-participants-2}

\resizebox{\textwidth}{!}
{\begin{tabular}{ |c|p{3.3cm}|p{1.6cm}|p{2.4cm}||p{2cm}|p{3.5cm}|p{3cm}|c| }
\hline
\rowcolor[HTML]{AAACED} \multicolumn{8}{|c|}{Number of Human Experts} \\
\hline
No.Years & Programming  Experience	& Industry Experience & Java Experience & Perform Maintenance & Perform Maintenance on Code not Authored  & Write Comments for the Code & Frequency \\\hline
10+ years & \centering\arraybackslash 2 & \centering\arraybackslash 0 & \centering\arraybackslash 1 & \centering\arraybackslash 0 & \centering\arraybackslash 0 & \centering\arraybackslash 4 & \centering\arraybackslash Daily \\
5-9 years & \centering\arraybackslash 10 & \centering\arraybackslash 7 & \centering\arraybackslash 9 & \centering\arraybackslash 3 & \centering\arraybackslash 1 & \centering\arraybackslash 3 & Weekly\\
1-4 years & \centering\arraybackslash 2 & \centering\arraybackslash 7 & \centering\arraybackslash 4 & \centering\arraybackslash 6 & \centering\arraybackslash 4 & \centering\arraybackslash 4 & Monthly\\
0-1 year & \centering\arraybackslash 0 & \centering\arraybackslash 0 & \centering\arraybackslash 0 & \centering\arraybackslash 5 & \centering\arraybackslash 9 & \centering\arraybackslash 3 & Yearly\\
\hline
\end{tabular}}
\end{table}

\paragraph{Metrics and Test.}

The first research question consists of three multiple choice questions which we assigned a values to the answers as higher value for stronger performance and preferred answer.
For the second research question we compared the keywords extracted by our approach and by the eye-tracking approach to the keywords chosen by the experts. Using this, we were able to compare the two approaches in terms of \emph{precision}, \emph{recall}, \emph{F-score}, and \emph{overall accuracy}, which are computed as follows:

\begin{equation}
\textit{Precision} = \frac{\#(\textit{RetrievedKeywords} \cap \textit{GoldSummaryKeywords})}{\#(\textit{RetrievedKeywords})}
\label{eq:eq_pre}
\end{equation}

\begin{equation}
\textit{Recall}  = \frac{\#(\textit{RetrievedKeywords} \cap \textit{GoldSummaryKeywords})}{\#(\textit{GoldSummaryKeywords})}.
\label{eq:eq_rec}
\end{equation}

\begin{equation}
\textit{F-Score}  =2 \times \frac{(\textit{Precision} \times \textit{Recall})}{(\textit{Precision}+\textit{Recall})}.
\label{eq:eq_fscor}
\end{equation}
In which $RetrievedKeywords$ are keywords extracted from the summary which is the output of the summarization approach and $GoldSummaryKeywords$ are keywords extracted from human expert's summaries.
Nenkova et~al.~\cite{nenkova} assert that recall is more effective in the evaluations of summaries than the precision. In the case using computing precision, perhaps some of the sentences returned by the approach are good enough, although they have not been chosen by the gold standard summary. Human judges often disagree on what the top n\% most important sentences are in a document. On the other hand, recall measures the overlap with already observed sentence choices. In other words, a precision value equal to one means that all the terms in the peer summary are relevant, though there could be relevant terms missing. On the other hand, a recall value equal to one means that the peer summary contains all the relevant terms, though it could also contain some irrelevant terms. Hence, there are some quality compromises between Precision and Recall that F-Score is used to handle these problems.

As Sridhara discussed in her paper~\cite{sridhara-towards}, there are two other factors in addition to the accuracy that must be computed for evaluating a generating summary approach; Content adequacy and conciseness. However we can equivalent these factors to precision and recall, respectively.

Finally, the \emph{overall accuracy}, that indicates the rate of correctly extracted keywords, and is computed as follows:
\begin{equation}
\textit{Overall Accuracy} = \frac{TP+TN}{TP+TN+FP+FN}
\label{eq:accuracy}
\end{equation}
In which,
\emph{TP} (\emph{True Positives}) is the number of keywords correctly extracted by the approach;
\emph{TN} (\emph{True Negatives}) is the number of keywords correctly not extracted by the approach;
\emph{FP} (\emph{False Positives}) is the number of keywords incorrectly extracted by the approach;
and \emph{FN} (\emph{False Negatives}) is the number of keywords incorrectly not extracted by the approach.

%%%%%%%%%%%%%%%%%%% RESULTS %%%
\subsubsection{Results}
This section presents the results of the evaluation of a user experiment in
which subjects are asked to write summaries for a set of methods and assess the quality of our generated summaries.
\paragraph{Quantitative Results.}

\begin{figure}
\begin{framed}
    \centering
    \subfloat[Experts' opinions about the accuracy of generated summaries.\label{subfig:accuracy}]{%
      \includegraphics[width=0.45\textwidth]{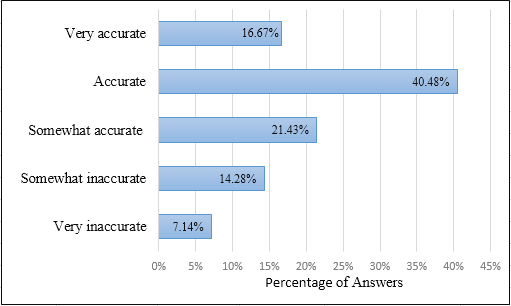}
    }
    \hfill
    \subfloat[Experts' opinions about the comprehensibility of generated summaries.\label{subfig:comprehensibility}]{%
      \includegraphics[width=0.45\textwidth]{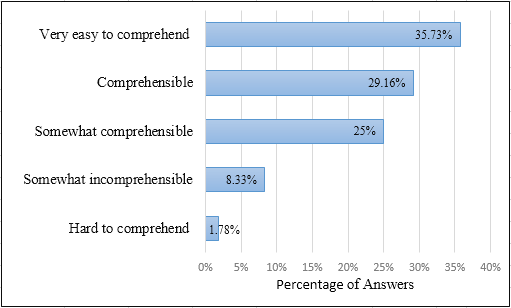}
    }
    \vfill
    \subfloat[Experts' satisfaction with the generated summaries.\label{subfig:satisfaction}]{%
      \includegraphics[width=0.45\textwidth]{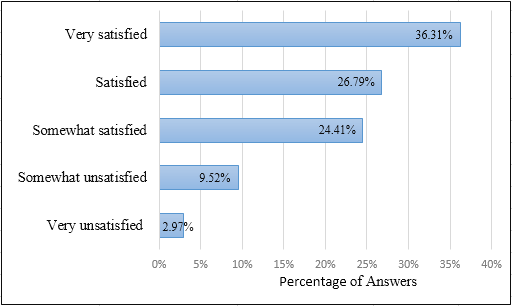}
    }
    \vfill
    \subfloat[The results of comparing CrowdSummarizer with the eye-tracking approach.\label{subfig:quantitative}]{%
      \includegraphics[width=0.65\textwidth]{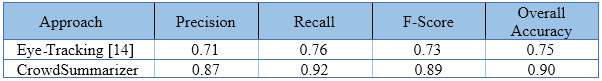}
    }
\end{framed}
\caption{The results of empirical evaluations of CrowdSummarizer.}
\label{fig:results}
\end{figure}

Figures~\ref{fig:results}\subref{subfig:accuracy}-\ref{fig:results}\subref{subfig:satisfaction} illustrate the results of empirical evaluations of the quality of CrowdSummarizer's automatically generated summaries. As can be seen in these figures, most of the experts participated in our study were satisfied with the quality of generated summaries in terms of accuracy and comprehensibility. Additionally, as indicated in \figref{results}\subref{subfig:quantitative}, CrowdSummarizer outperforms the eye-tracking approach in terms of precision, recall, F-Score, and overall accuracy. This finding indicates that the different areas in the different kinds of methods must have different weights according to the users' priorities.

\paragraph{Qualitative Results.}

This section reports participants' opinions about automatically generated summaries.

\textit{''The automated summaries are well defined for simple and illustrative methods. the sentences are all intelligently produced and the length of summery is appropriate. but it should be noted that for big and complicated methods the sentences are not such intelligently made and we see the actual variables and code lines in summary while we expect to see more user friendly words and shorter sentences in summary text."}

\textit{''The summaries explained the function parameters and variables  well, but explaining the purpose of the methods can be improved."}

\textit{''The summarizer produces short explanations about each method. The summary's length seems adequate to me, moreover the input and output declarations are understandable. Also the text defines each function call clearly which helped me understand the relations between several functions. However I did not feel comfortable about condition description template, it is so much similar to the code."}
%%%%%%%%%%%%%%%%%%%%%%%%%%%%%%%%%%%%%%%%%%%
%%%
%%%             Threats to Validity
%%%
%%%%%%%%%%%%%%%%%%%%%%%%%%%%%%%%%%%%%%%%%%%
\subsection{Threats to Validity}
There are several factors that may potentially affect the validity of the results of this experiment. This section provides a description of these factors.
\paragraph{Internal Validity.}
The main threat in this part is the distribution of participants over the methods. We addressed this threat by assigning methods from different programs to participants randomly. As the second point, implementation of SWUM is another factor which influence on the generated summaries.

\paragraph{External Validity}
The main threat to external validity concerns the programs and the methods had been selected and it may be deducted that our summariztion approach is prominent for a set of determined methods. We minimized this threat by selecting 78 methods from 11 applications across different domains and different sizes. Also, stress, human errors and fatigue are effective factors that decrease reliability of the results. We addressed this threat by selecting the great number of participants from a diverse range of experience with at least 1 years of experience.

\paragraph{Construct Validity}
The main source for a threat to construct validity is that we need well-defined measures to compare the effectiveness of different source code summarization approaches. We used
the standard measures used in the summarization literature~\cite{rodeghero}~\cite{sridhara-towards}, such as \emph{Precision}, \emph{Recall}, \emph{Accuracy}, \emph{Content Adequacy} and \emph{Conciseness}.
All the summaries used in the second study are generated by the CrowdSummarizer tool for the methods from the open-source Java applications used in the other approaches.

\paragraph{Replicability}
The natural language templates used generating summaries, the data collection, the experimental materials, including the subjects' background questionnaires, experiment questionnaires are available online~\cite{supportingmaterials}.
%%%%%%%%%%%%%%%%%%%%%%%%%%%%%%%%%%%%%%%%%%%
%%%
%%%             Discussion
%%%
%%%%%%%%%%%%%%%%%%%%%%%%%%%%%%%%%%%%%%%%%%%
\section{Discussion}\label{discussion}
Here, we explain our experiences and lessons learned during the crowdsourcing process.
\subsection{Length of Summaries}\label{length}
As evidenced by the results, for the 4000 summary in initial examination of the CrowdSummarizer, developers select the number of keywords in their summaries based on the method complexity, not just length of the method. As an example, they used 6 keywords(as average) in their summaries for a method with 41 keywords, and 7 keywords(as average) for a method with 242 keywords. Meanwhile, 14 keywords is an average for the number of keywords in the summaries for a method with length of 87 keywords (Table \ref{tab:lenght_method}).

\begin{table}[h]
	\centering
	\small
    \caption{The correlation between method lengths and summary lengths for 20 randomly selected methods.}
    \label{tab:lenght_method}
\newcolumntype{s}{>{\columncolor[HTML]{AAACED}} c}
\definecolor{green}{RGB}{112, 173, 71}
\resizebox{\textwidth}{!}
{\begin{tabular}{ |c|s||c|s||c|s| }
\hline
Method Length & Average of Summary Lengths & Method Length & Average of Summary Lengths & Method Length & Average of Summary Lengths  \\
\hline\hline
69 & 10.21 & 62 & 12.7 & 41 & 5.42\\
\hline
92 & 8.88 & 124 & 10.69 & 90 & 11.9\\
\hline
67 & 11.32 & 220 & 10.43 & 113 & 9.44\\
\hline
74 & 6.63 & 242 & 7.85 & 54 & 10.13\\
\hline
76 & 11.64 & 13 & 11.53 & 123 & 13.07\\
\hline
89 & 5.76 & 70 & 4.12 & 87 & 14.921\\
\hline
31 & 9 & 46 & 7.41 & \cellcolor{green} Avg = 95.65 & \cellcolor{green}Avg = 8.74\\
\hline
\end{tabular}}
\end{table}

The main difference that affects on the length of summaries of the method with 87 keywords and the method with 242 keywords is in conditional statements that the former has which includes three \textit{for} loops, one \textit{while} loop and two \textit{switch} statements.
\subsection{The Summary and the Method Overlap}
Specially, We computed a factor \textit{Summary and Method Overlap(SMO)} as follows:
\begin{equation}
SMO(s,m) = \frac{|SummaryKeywords(s) \cap (MethodKeywords(m) \cup Syn(MethodKeywords(m)))|}{|SummaryKeywords(s) \cup (MethodKeywords(m) \cup Syn(MethodKeywords(m)))|}
\label{eq:eq_smo}
\end{equation}

Where $SummryKeywords(m)$, $MethodKeywords(m)$ and $Syn(MethodKeywords(m))$ are the sets of the terms extracted
from the summary $s$, method $m$ and synonym set of method $m$, respectively.
The measure has a value between [0, 1], that a higher value shows more use of method keywords in summaries. Indeed, based on~\cite{mcburney-empirical}, we expect that if a summaries get a higher $SMO$ value it will get more acceptance from the developers.
We observed that, methods with more than 2 parameters and methods with the highest fan-out have the highest $SMO$ values.
In addition, Table \ref{tab:smo} shows that, although, people with less programming experience use more method keywords in their summaries, but there is no obvious relationship between $SMO$ measure and developer's programming experience(i.e. Programmers in the all levels of experience use the method words in their summaries with the close rate.)

\begin{table}[h]
	\centering
	\small
    \caption{$SMO$ measure for different players' programming experience.}
    \label{tab:smo}
    \definecolor{green}{RGB}{112, 173, 71}
{\begin{tabular}{ |c|c| }
\hline
\rowcolor[HTML]{AAACED} No.Months of Programming Experience & Average of SMO\\
\hline\hline
84+ months & 0.176187244\\
\hline
48-84 months & 0.197456277\\
\hline
24-48 months & 0.225091266\\
\hline
6-24 months & 0.19564788\\
\hline
0-6 months & 0.248784946\\
\hline
\rowcolor{green} Average & 0.20\\
\end{tabular}}
\end{table}
%%%%%%%%%%%%%%%%%%%%%%%%%%%%%%%%%%%%%%%%%%%
%%%
%%%             Related Works
%%%
%%%%%%%%%%%%%%%%%%%%%%%%%%%%%%%%%%%%%%%%%%%
\section{Related Work}\label{sec:relatedwork}
This section will cover the related works on source code summarization as well as code documentation using the crowdsourcing concept.
\subsection{Source Code Summarization}
The most related work to CrowdSummarizer is the eye-tracking approach~\cite{rodeghero} which is an improvement of the technique proposed by Haiduc et~al.~\cite{haiduc-use} for extracting the most relevant keywords from a method by applying the text retrieval technique VSM\footnote{Vector Space Model} tf-idf. This approach uses an eye-tracking system to get data about the keywords that programmers view as important while summarizing a method. However, this approach has some drawbacks like the limited length for the methods it can summarize, the few numbers of people who participated in its study, and fatigue which affects the eye tracking results.
G. Sridhara~\cite{sridhara-automatic-generation:PhD:2012} uses a number of heuristics in his PhD thesis to automatically generate a natural language description about just those program statements of a method that implement a particular behavior. Thus, unlike our approach, it does not generate summaries for the whole method. Rastkar et~al.~\cite{rastkar-generating} proposed a different approach to improve productivity of developers in maintenance phase by generating summaries for cross-cutting concern in source code. The summaries describe parts of a concern and its interaction with other concerns. Moreno et~al.~\cite{moreno-automatic} presented a different approach to generate natural language summaries for Java classes based on the class's stereotype and heuristics. The purpose of this approach is to give information about the class's responsibilities and not the relationship with other classes. A similar work based on stereotypes and static analysis is introduced by Abid et~al.~\cite{abid2015using} for generating summaries for C++ methods. The methods stereotypes in this approach are identified by the approach~\cite{Dragan}.

Describing high level actions using topic modelling approaches are also considered in~\cite{eddy-eval} and~\cite{mcburney-improving}. Eddy et~al.~\cite{eddy-eval} replicated and expanded the approach proposed by Haiduc et~al.~\cite{haiduc-use} by introducing Hirarchical Pachinko Allocation Model(hPAM), as a new topic modelling technique. McBurney et~al.~\cite{mcburney-improving} proposed an approach to extract topics in source code using program's call graph and a topic modelling approach named Hierarchical Document Topic Model (HDTM).
A recent work in~\cite{McBurney:TSE:2016} helps programmers understand the role a method plays in a program. In particular, it summarizes the context of a method, like how it is called or its output is used. Consequently, it does not summarize the method itself. There are other works which generate descriptions for other software artifacts, such as bug reports~\cite{rastkar-summarizingSoftware} and execution traces~\cite{hamou2006summarizing}.

We tried to tackle the restrictions of existing approaches by using the power of the crowd as the source of valuable information to generate summaries, as well as using the LDA technique to extract latent topics in a program.
\subsection{Crowd Documentation}
Crowdsourcing has attracted significant recent attentions to support a wide range of software engineering activities, such as requirement, design, coding, testing and evolution and maintenance. Here, we briefly discussed crowd software documentation that is the closest to our work. Jiau and Yang~\cite{jiau} enhanced inequality of documentation based on object oriented inheritance using documentation reuse. Conducting an empirical study on the three Java APIs in Stack Overflow showed their approach feasibility. Pawlik et~al.~\cite{pawlik} conducted a case study using a python library for scientific computing called Numpy. The case study highlighted benefits of using crowdsourcing in software documentation as well as considerations in planning, organizing and implementing crowdsourcing for software documentation. Parnin et~al.~\cite{parnin} showed that question and answer websites, such as Stack Overflow can be used for creating software documentation by using API functionality discussions and code examples. As it turns out, Stack Overflow is one of the best sources for creating code documentation, however, no consensus has developed around generating a natural language high quality summary for a method.

The key difference between our approach (as well as other summarization approaches) and these existing approaches is that documentation techniques, such as JavaDocs~\cite{kramer} work on combining short paragraphs attached to each method to create a code document and ignore the way that this short paragraphs are created (manual (i.e. human written) or automatically), while code summarization techniques generate descriptions about a piece of a code (i.e. the short paragraph itself). Indeed, the code summaries can be considered as the main components of the program document.

In addition, one of the challenges of evaluating code summarization techniques is the cost
of  the  empirical  studies  involving  users  assessing  different  intrinsic quality  attributes  of summaries. the Crowdsourced-based  summaries  represent  a  good starting  point  to  build a benchmark or gold set that allows to perform a quantitative, intrinsic assessment of those quality attributes would facilitate and reduce the cost of of evaluating code  summarization  techniques
%%%%%%%%%%%%%%%%%%%%%%%%%%%%%%%%%%%%%%%%%%%
%%%
%%%             Conclusions
%%%
%%%%%%%%%%%%%%%%%%%%%%%%%%%%%%%%%%%%%%%%%%%
\section{Conclusions and Future Work}\label{sec:conclusion}
This article introduced CrowdSummarizer, a novel approach for automated generation of natural language summaries for Java programs and their methods. To this end, CrowdSummarizer employs the concepts of crowdsourcing, gamification, and natural language processing. Furthermore, it applies the LDA technique to identify the latent topics in a Java program, and to classify its methods into those topics. We have implemented CrowdSummarizer as an Eclipse plugin that works with a web-based code summarization game that can be played by the crowd. The results of an empirical study conducted by 149 developers with different levels of experience illustrated that CrowdSummarizer is applicable and motivating for developers to use it in practice. Moreover, the results of another empirical study done with the help of 14 experts indicated that developers are satisfied with the quality of automatically generated summaries in terms of accuracy and comprehensibility.
In future, CrowdSummarizer can be easily extended to include other programming languages like C++ as well. In addition, we plan to work on other text retrieval techniques to extract keywords from the method. We also intend to generate natural language summaries for classes using the crowd and the CrowdSummarizer. Finally, we want to use dynamic analysis in combination of static and linguistic information.

%%%%%%%%%%%%%%%%%%%%%%%%%%%%%%%%%%%%%%%%%%%
%%%
%%%             Acknowledgement
%%%
%%%%%%%%%%%%%%%%%%%%%%%%%%%%%%%%%%%%%%%%%%%
\section{Acknowledgement}
The authors would like to thank Dr. Dave Binkley for providing identifier splitting. Finally, They would also like thank the 149 participant who spent time for assessing the CrowdSummarizer platform and all the 14 subjects who participated in the user study experiment.
%%%%%%%%%%%%%%%%%%%%%%%%%%%%%%%%%%%%%%%%%%%
%%%
%%%             Bibliography
%%%
%%%%%%%%%%%%%%%%%%%%%%%%%%%%%%%%%%%%%%%%%%%
%\bibliographystyle{ieeetr}
%\bibliography{TechReport}

%%%%%%%%%%%%%%%%%%%%%%%%%%%%%%%%%%%%%%%%%%%
%%%
%%%          About the Authors
%%%
%%%%%%%%%%%%%%%%%%%%%%%%%%%%%%%%%%%%%%%%%%%
\vspace{1cm}
\noindent{\Large\textbf{About the Authors}}

\paragraph{Sahar Badihi} is a master's student at the Sharif University of Technology. Her research interests include software engineering, program comprehension, and code summarization. Contact her at \url{sbadihi@ce.sharif.edu}.

\paragraph{Abbas Heydarnoori} is an assistant professor at the Sharif University of Technology. Before, he was a post-doctoral fellow at the University of Lugano, Switzerland. Abbas holds a PhD from the University of Waterloo, Canada. His research interests focus on software evolution, mining software repositories, and recommendation systems in software engineering. Contact him at \url{heydarnoori@sharif.edu}.

\end{document}